\documentclass[preprint,12pt]{elsarticle}
 \usepackage{hyperref}
\usepackage{amssymb}
\usepackage{graphicx}
\graphicspath{ {Figures/} }
\usepackage{amsmath}
\usepackage{mathtools}
\usepackage{setspace}
\usepackage{array}
\usepackage{ragged2e}
\usepackage[ruled,lined]{algorithm2e}
\usepackage{tikz}
\usepackage{float}
\usepackage{subfigure}
\usepackage{cleveref}
 
\usepackage{booktabs}

\journal{Knowledge-Based Systems}

\newdefinition{rmk}{Remark}
\newproof{pf}{Proof}
\newdefinition{defi}{Definition}
\begin{document}
	
	\begin{frontmatter}
	%
\title{Deep Learning Approach on Information Diffusion in Heterogeneous Networks}
	\author[rv1]{ Soheila Molaei}
	\author[rv1]{Hadi Zare\corref{cor1}}
	\author[rv1]{Hadi Veisi}
	\cortext[cor1]{Corresponding author}
	\address[rv1]{Faculty of New Sciences and Technologies, University of Tehran, Tehran, Iran.}
	
\begin{abstract}
There are many real-world complex systems with multi-type interacting entities that can be regarded as heterogeneous networks including  human connections and biological evolutions. One of the main issues in such networks is to predict information diffusion such as shape, growth and size of social events and evolutions in the future. While there exist a variety of works on this topic mainly using a threshold-based approach, they suffer from the local viewpoint on the network and sensitivity to the threshold parameters.
In this paper, information diffusion is considered through a latent representation learning of the heterogeneous networks  to encode in a deep learning model. 
To this end, we propose a novel meta-path representation learning approach, Heterogeneous Deep Diffusion(HDD), to exploit meta-paths as main entities in networks.
At first, the functional  heterogeneous  structures of the network are learned  by a continuous latent representation through traversing meta-paths with the aim of  global end-to-end viewpoint.
Then, the well-known deep learning architectures  are employed on our generated features to predict diffusion processes in the network. The proposed approach enables us to apply it on different information diffusion tasks such as topic diffusion and cascade prediction. 
We demonstrate the proposed approach on benchmark network datasets through the well-known evaluation measures. The experimental results show that our approach outperforms the earlier state-of-the-art methods.
\end{abstract}
\end{frontmatter}

\section{Introduction}
Information diffusion is one of the widely studied dynamical processed on networks. Information such as news, innovations, and viruses start from a set of seed nodes and propagates throughout the network\cite{bakshy2012role,zhang2016dynamics}. Information diffusion has been investigated in a wide range of fields including health care \cite{Hu2017,Green2002}, complex networks \cite{Kirst2016, Zhang2016} and social networks \cite{Dhamal2016, Margaris2016}.
One of the most important tasks on networked systems is understanding,  modeling\textcolor{black}{,} and predicting the rapid events and evolutions in the body of \textcolor{black}{a} network. This is mainly motivated by the well-known fact that discovering the structure of networks is resulted to predict the patterns of social events such as their shape, size\textcolor{black}{,} and growth known as information diffusion \cite{li2017deepcas}. Many researchers have investigated various techniques and approaches to model information diffusion on homogeneous and heterogeneous networks such as \cite{Gui2014, Yang2015, wang2017model,nian2010efficient}. \par
Formally, an information network denoted as a graph $G=(V,E)$ with the set $V$  of nodes and set $E$ of edges, is homogeneous if and only if the edges and nodes are of the same type, and heterogeneous if different types of nodes and  relations on edges are involved \cite{Sun2013a}.
Various studies have been conducted on homogeneous networks including semantic parsing \cite{bordes2012joint}, control of epidemic diseases \textcolor{black}{\cite{wang2016model,ren2014epidemic} } and link prediction \cite{Bordes2013, Socher2013, Wang2014}. Recently, heterogeneous networks have been considered as an attractive research field, due to the fact of more natural assumption of heterogeneous networks on many real-world phenomena\textcolor{black}{\cite{liu2010inhomogeneity,wen2013inhomogeneity}}. 
Watts \cite{Watts2002} studied the role of threshold values and network structure in information diffusion.
Therefore, information diffusion, as a particular topic of interest in this regard, has been studied in \cite{Gui2014} using different meta-paths in heterogeneous networks. In  \cite{Gui2014}, MLTM-R is proposed by distinguishing the power in passing information around for different types of meta-paths. 
Pathsim was considered as a weight between each two nodes in this method through which predictions were conducted \cite{Kuck2015,shang2016meta}.\par
Recently, much attention has been focused on deep learning in heterogeneous networks \cite{audebert2017fusion,liu2016deep}.  In \cite{wu2017coupled}, a deep learning framework is investigated as coupled with deep learning (CDL), to address the VIS-NIR heterogeneous matching problem via the topic diffusion on networks. Tang et al. \cite{tang2015line}  proposed the LINE algorithm to learn embedding which traverses all edge types and samples one edge at a time for each edge type. Chang et al. \cite{chang2015heterogeneous} proposed a deep architecture by embedding Heterogeneous Information Network(HIN). \textcolor{black}{metapath2vec \cite{dong2017metapath2vec} was proposed as a} new HIN algorithm which transcribes semantics in HINs by meta-paths. 
With this in mind, we point out the strengths and weaknesses of the existing approach to topic diffusion in heterogeneous networks. Our motivation is to employ a deep learning approach on information diffusion tasks such as topic diffusion and cascade prediction to alleviate the problems of the earlier ones. Most  works on topic diffusion suffer from local similarity considerations and single node-based embedding \cite{Gui2014}, our latent based representational learning approach investigates a global view of the activation prediction modeling. 
\textcolor{black}{ Local similarity computations for each two nodes are time-consuming and unendurable for big networks. While there are various similarity measures,  the choosing of  appropriate similarity method is a challenge.}

Here, we propose a novel latent representational learning approach on heterogeneous networks based on different types of meta-paths. \textcolor{black}{Due to the challenges of similarity calculation, we omit this step and for representing a node we consider different graph meta-paths which in each meta-path graph, the neighbors of a node could be different.}
Furthermore, the proposed approach is employed on different deep neural network architectures to predict information diffusion tasks in an end-to-end framework. The experimental results demonstrate the strength of the proposed approach as compared with the earlier works. We evaluate the performance of the proposed method using real-world information graph i.e. DBLP, PubMed\textcolor{black}{, ACM, APS, and Citeseer}. HDD is compared with multiple strong baselines, including feature-based methods \textcolor{black}{such as deepwalk \cite{perozzi2014deepwalk}, } node2vec \cite{grover2016node2vec}, deepcas \cite{li2017deepcas}, and MLTM-R \cite{Gui2014}. HDD method significantly improves the results over these baselines.

\textcolor{black}{In this paper, we have applied CNN-LSTM to generate  meta-paths representation for information diffusion tasks. The CNN-LSTM architecture contains a CNN layer for feature extraction along with an LSTM to support sequential prediction. This model draws on the intuition that the sequence of features extracted from CNN can be encoded into a vector representation using LSTM architecture.  CNN-LSTM can embed the the whole meta-paths in the network. Our representations are general-purpose which can be applied on heterogeneous network for information diffusion and cascade prediction tasks. To summarize, our work makes the following contributions,
\begin{itemize}
		\item We propose a Heterogeneous Deep Diffusion (HDD) model to learn HIN node embeddings which can be retrieved for downstream tasks,  such as topic diffusion and information cascade.
		\item We design a feature extraction mechanism to conduct weighted aggregations of neighborhood nodes on different meta-paths for learning comprehensive embeddings.
		\item We conduct experiments on real-world datasets to show the superiority of our model against the prior state-of-the-art methods and give a comprehensive analysis of the learned embeddings in order to gain more insights from the datasets.		
\end{itemize}
}
The remainder of this paper is organized as follows. In Section \ref{ssec:Literature}  related works are reviewed on the heterogeneous networks along with a brief motivation for the idea. Section \ref{ssec:Methods} describes the proposed deep learning framework of information diffusion for heterogeneous networks. Section \ref{ssec:Results} is devoted to the experimental settings and results on the benchmark real networks. Finally, we conclude the paper in Section \ref{ssec:Conclusion} with some suggestions for further works.
\section{Literature Review}\label{ssec:Literature}
There exist a multitude of works on exploiting heterogeneous networks to uncover the structural patterns by considering the rich side of information on different nodes objects and edge attributes in these networks \cite{shi2017survey}. Sentiment classification of product reviews using heterogeneous networks of users, products, and words was addressed by Zhou et al. \cite{Zhou2007}. In this regard, Zhou et al. \cite{Zhou2007} proposed a co-ranking method which classifies the authors and documents separately based on random walks.
Angelova et al. \cite{Angelova2012} presented a new classification method for mining of homogeneous information networks through their decomposition into multiple homogeneous ones. The idea of citation recommendation using heterogeneous networks was proposed by Liu et al.\cite{Liu2014}. Information diffusion has been used in such networks. Information diffusion is mainly fallen into two categories as topic diffusion and cascade prediction which are described as follows:

\subsection{Topic Diffusion}
\textcolor{black}{There is} a variety of work on topic diffusion as a primary task \textcolor{black}{to analyze} the heterogeneous networks \textcolor{black}{that have been} employed in applied domains such as medical and public health issues. Several works considered the epidemic modeling in heterogeneous networks like infections spreading on population systems \cite{Moreno2002}, modified SIR model \cite{Yang2007}, and a survey \cite{salehi2015spreading}. Wang and Dai \cite{Wang2008} addressed virus spreading in heterogeneous networks by applying an epidemic threshold on the well-known SIS model.  \cite{Yang2015}  showed leveraging a heterogeneous network among people to yield more resistance against the epidemic spread of the virus. Epidemic spreading is an important issue that was considered in other networks likes time-varying networks \cite{nadini2017epidemic} and adaptive ones \cite{demirel2017dynamics}. Nadini et al. \cite{nadini2017epidemic} used SIR and SIS models and investigated effects of modular and temporal connectivity patterns on epidemic spreading.\par
Various techniques are proposed on topic diffusion in heterogeneous networks. The degree distribution is used in \cite{Sermpezis2013},  for the modeling of information diffusion by taking the assumption of diffusion between two nodes at random times. Zhou and Liu \cite{zhou2013social} presented a social influence based clustering framework. Molaei et al. \cite{molaei2018information} predicted topic diffusion process in heterogeneous networks by considering the interactions of different meta-paths.
A heterogeneous network based model was proposed for new products diffusion in two stages framework \cite{Li2014}.
In \cite{Boccaletti2014}, the concept of heterogeneous networks was used as an alternative definition for the infrastructure networks to explain the diffusion process. \textcolor{black}{In contrast to most prior approaches to topic diffusion in multilayer networks with probabilistic graphical models, we have used deep representation learning which automatically extracts summarized features from each node in such networks.}

\subsection{Cascade Prediction}
There are many cascade prediction methods originating from different research areas as both classification and regression problems. Recently, methods based on representation learning emerge with impressive predictive power. Graph representation learning methods have largely been based on the popular skip-gram model \cite{mikolov2013efficient, cheng2006n} originally introduced for learning vector representations of words in the text. In particular, DeepWalk \cite{perozzi2014deepwalk} used this approach to embed the nodes such that the co-occurrence frequencies of pairs in short random walks are preserved. Node2vec \cite{grover2016node2vec} introduced hyperparameters to DeepWalk that tune the depth and breadth of the random walks.
\par
Lately, some deep learning based models show good performances. These models learn to predict information cascade in an end-to-end manner. DeepCas \cite{li2017deepcas} uses random walk to sample paths from different snapshots of the graph then uses Gated Recurrent Units (GRU) and attention mechanism to extract features from random walk paths to predict information cascade. DeepHawkes \cite{cao2017deephawkes} uses GRU to encode each cascade path, and employs weighted average pooling based on time decay effect to combine features from all cascade paths.\par 
While there exist some recent works on cascade predictions with a deep learning approach \cite{li2017deepcas, cao2017deephawkes}, our proposed framework differs from them due to the type of used networks, applied methods, input parameters and also more generalization capability to the different information diffusion process. We focus on the heterogeneous network and we added weights to the input with considering different meta-paths. The deep learning approach exploits the representational learning on meta-paths to generate the required features. \textcolor{black}{On the other hand, we extracted features from networks differently which used more information and considered more relation in such networks.}

\section{Proposed Method}\label{ssec:Methods}
\begin{figure}[H]
	\centering
	\includegraphics[width=1\linewidth]{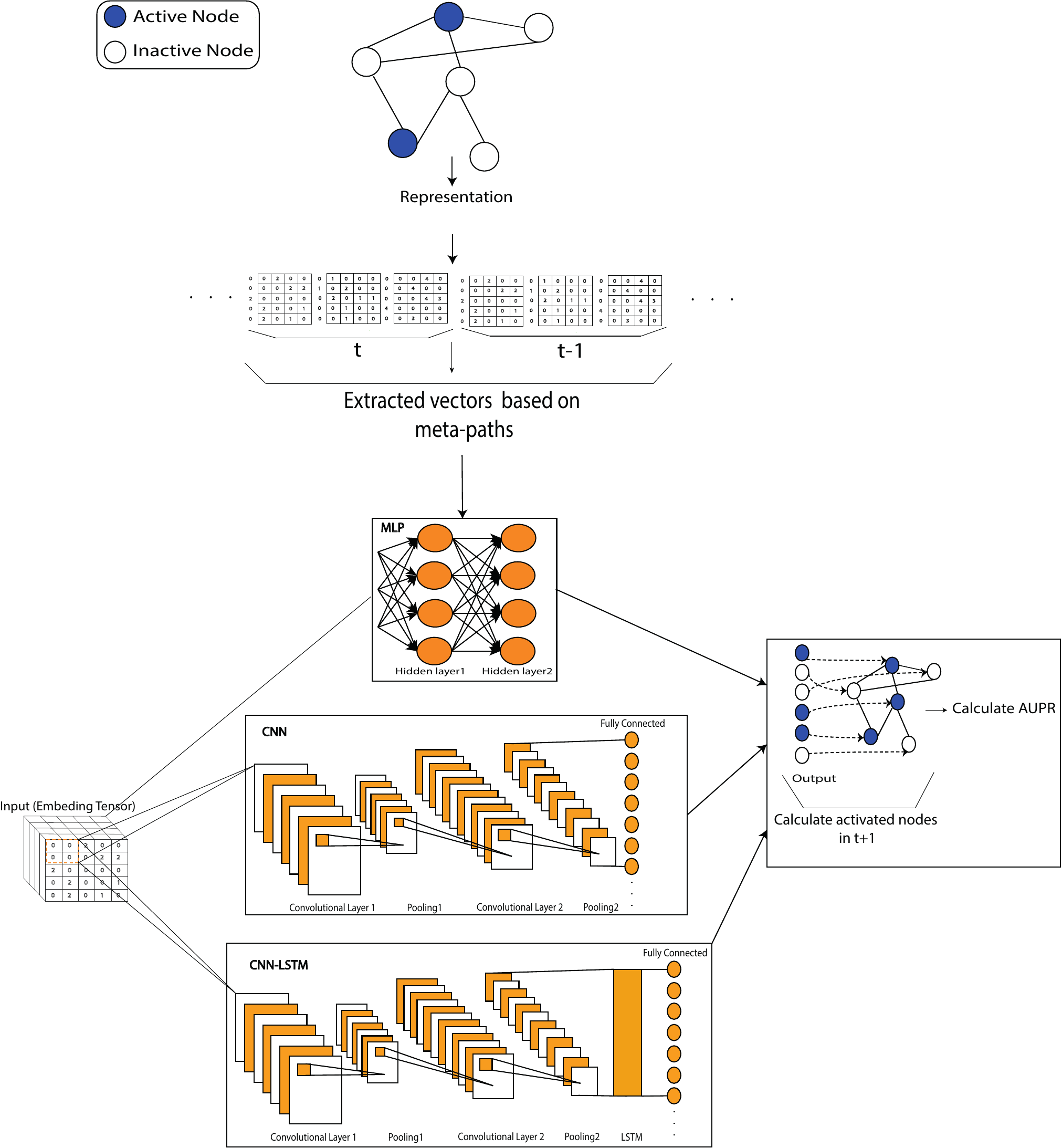}
	\caption{The overall procedure of the proposed method}
	\label{fig:generalschemepaper}
\end{figure}

We present a general framework, HDD(Heterogeneous Deep Diffusion) on topic diffusion in heterogeneous networks through a deep learning approach.  The flow-graph of the overall structure of the proposed approach is shown in Figure \ref{fig:generalschemepaper}. 

\begin{figure}[H]
	\centering
	\includegraphics[width=0.7\linewidth]{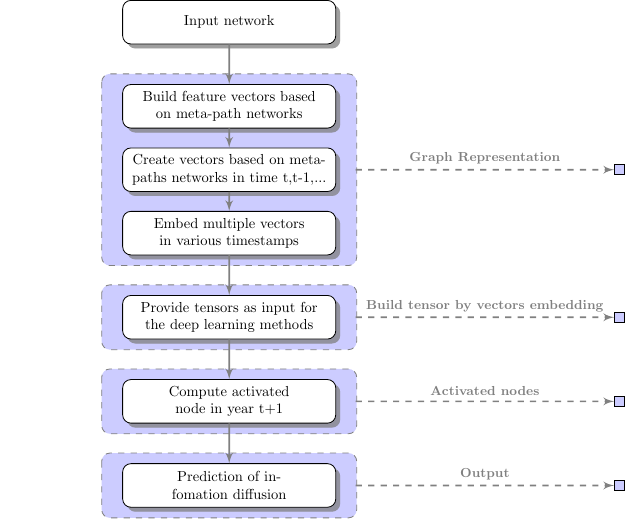}
	\caption{The overall flowchart of the proposed method}
	\label{fig:generalschemepaper1}
\end{figure}

Initially, the graph with distinct node types as active and inactive ones is given. In the representation stage, the continuous latent features are learned based on different meta-path definitions. Therefore,  each node is represented as a continuous vector in a sequence of different meta-paths.  We employ a variety of deep neural networks architectures on sequences of nodes representations to predict outgoing activation or inactivation in the body of  network. 
In following, we describe the main information diffusion tasks in our formulation and then present our approach. \textcolor{black}{The overall flowchart of the proposed method is mentioned in Figure \ref{fig:generalschemepaper1}.}
\textcolor{black}{For each meta-path we create a meta-path graph and then update the embedded information of each node in these graphs within a vector. As Figure 3 shows on APA meta-path, a vector is generated for each node, in which the cells are the number of papers two nodes have co-authored on a specified topic in the year t. The last column is dedicated to output which is 1 (active node) if in t + 1 two nodes will co-author a paper on a specific topic, otherwise considered as 0 (inactive node). Finally, the creation of these matrices in multiple times through a variety of meta-paths yields to a tensor as the input on our selected models for downstream prediction tasks as illustrated in Figure \ref{fig:generalschemepaper}.}
	
\begin{figure}[H]
	\centering
	\includegraphics[width=1\linewidth]{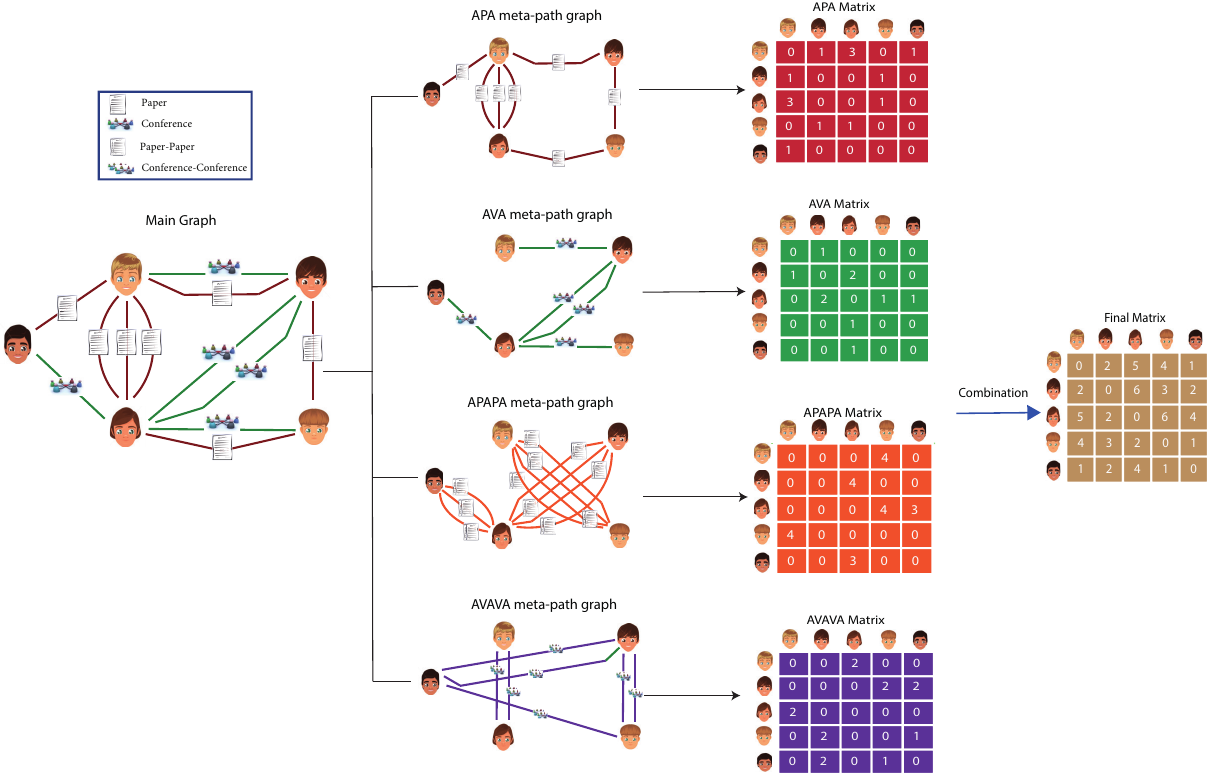}
	\caption{The procedure of a tensor input creation in a time-stamp \textit{t}}
	\label{fig:procedure}
\end{figure}

\subsection{ Problem  Statement}
The two main tasks of information diffusion,  \emph{topic diffusion} and \emph{cascade prediction} are described  on heterogeneous network setting.  
\subsubsection{Topic Diffusion}
In a general scheme of network \(T_G=(A,G)\), meta-path $P$ is defined where \(A\) and \(R\) represent the type of nodes and edges (meta-path). It is displayed as  \(A_1R_1A_2R_2...R_lA_{l+1}\). 
\begin{figure}[H]
	\centering
	\includegraphics[width=0.6\linewidth]{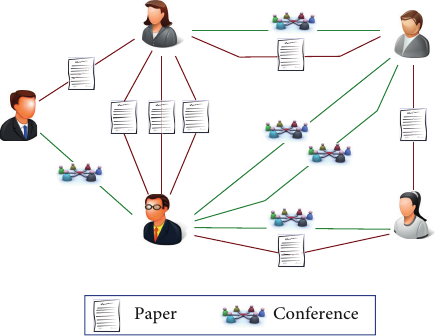}
	\caption{DBLP network}
	\label{fig:figure1}
\end{figure}
A scenario of heterogeneous collaboration network is represented in Figure \ref{fig:figure1} where an edge between two authors denotes a common publication. There may be multiple edges between two nodes for authors with multiple publications together. The “APA” presents the meta-path of coauthors on a paper (P) between two authors (A), and in the same way, “AVA” denotes authors (A) publish papers in the same conference (V). 

\par
Topic diffusion aims to solve the problem of who will write a paper on a particular topic at the time $t + 1$, when a node $i$ wrote a paper on the same topic at time $t$.

%
Here, our aim is to construct a universal framework to employ the meta-path objects (instances) based on deep learning ideas to alleviate the learning of $\beta_k$ computations in the prior works. 

\subsubsection{Cascade Prediction}
\begin{figure}[H]
	\centering
	\includegraphics[width=0.7\linewidth]{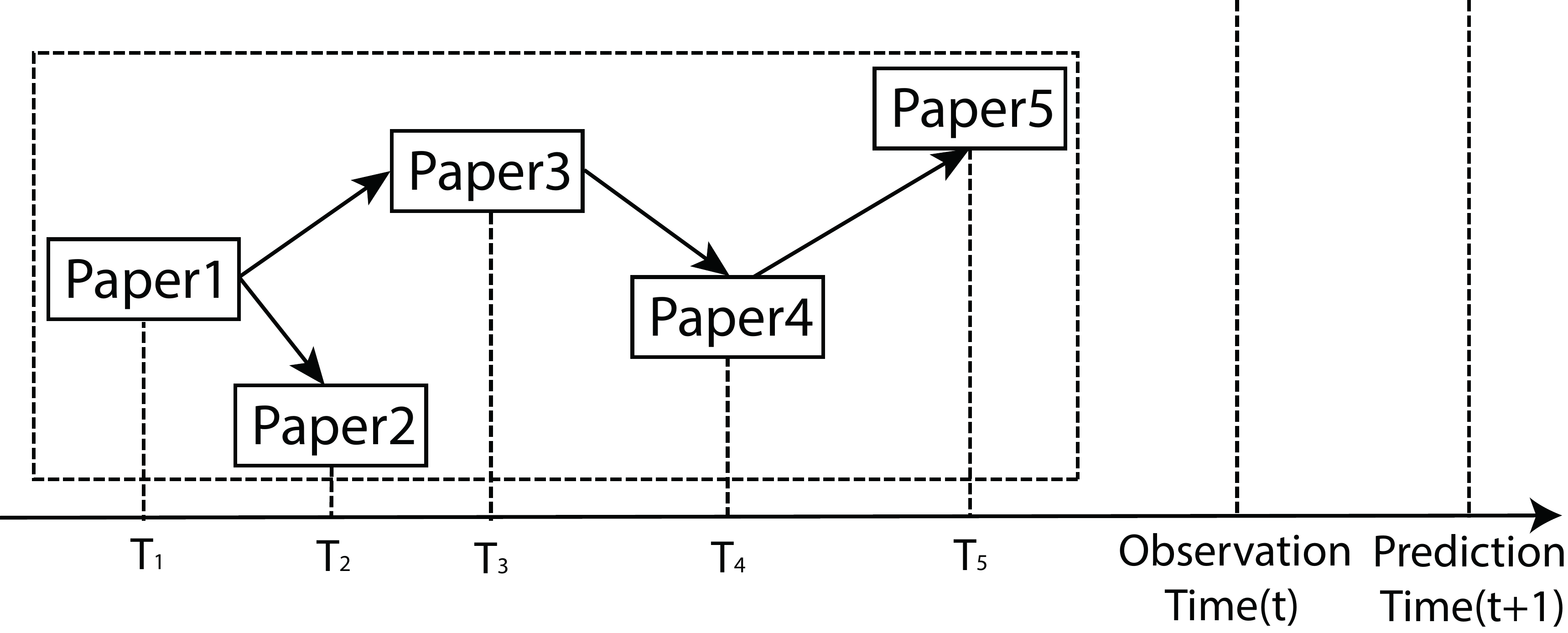}
	\caption{Cascade definition}
	\label{fig:cascad2}
\end{figure}

For constructing cascades in the graph $G=(V,E)$, citation relationship between nodes $i$ and $j$ is required.
The cascade path is defined according to the reference of its citing papers. Suppose we have $N$ topics denoted by $N={n_i}(1<i<N)$. For each topic $n_i$, we use a cascade $C_i= {(u_i,v_i,t_i)}$ to record the topic diffusion process of $n_i$ meaning that the author $v_i$ cites paper of the author $u_i$ and the time elapsed between the citation process is $t_i$. Figure \ref{fig:cascad2} represents the cascades which are denoted by $(Paper1, t_1 =0)$, $(Paper1, Paper2, t_2)$, $(Paper1 ,Paper3, t_3)$, $(Paper3, Paper4, t_4)$, $(Paper4, Paper5, t_5)$.
 In our framework, the aim is to predict subsequent cascades of the specific topic at  time $t+1$  from the current $t$-th snapshot graph.
 

\subsection{Heterogeneous Deep Diffusion}
To model topic diffusion in the heterogeneous networks, we introduce Heterogeneous Deep Diffusion(HDD) method by exploiting the meta-paths. In this part,  meta-path graphs are extracted from the raw datasets to represent nodes in a heterogeneous network $G=(V,E)$.
 A snapshot of graph $G$ at time $t$ is characterized by a meta-path graph $G_t^k=(V^k_t,E^k_t)$, where $V^k_t$ is a subset of nodes in $V$ that have adopted meta-path $k$ at time $t$. 
In our setting,  graph snapshots are considered on different timestamps along with a variety of meta-path definitions.


 \subsubsection{Author Embedding }
 Each node is represented as a vector, $q \in R^ N$ where $N$ is the total number of authors. All users share an embedding tensor $M \in R ^{N*N*T_s}$, which $T_s$ is the number of timestamps as depicted in Figure \ref{fig:generalschemepaper}.

\subsubsection{Meta-path Encoding }
\textcolor{black}{The proposed HDD model extracts the semantic relationships among nodes in the semantic space based on different meta-paths and creates a meta-path graph for each individual meta-path; thus, the adjacency matrix in each meta-path graph is different. After extracting neighborhood features from each meta-path graph, at the final step, the extracted features are embedded together.}

To represent the information flow, we use a variety of neural network architectures like LSTM, CNN, and CNN-LSTM. The main key factor here is considering all meta-paths. All of the meta-paths author embedding tensors are merged to a unique  tensor in our framework at different timestamps rather than applying ad-hoc based relationship. 
The meta-paths embedding in a unique tensor enriches the proposed approach from the following perspectives, 
\begin{enumerate}
\item \textbf{Loss of information:} In the real world, two authors may have different relationships to be considered rather than single relation to prevent the loss of significant information. 
\item  \textbf{Time sequences:} Different snapshots of the meta-path graphs $G^k$ are yielded to distinct author embedding cascades over time. 
\end{enumerate}
\par
We encode the entire met-paths for each author by using embedding tensor as an input through an LSTM \cite{hochreiter1997long}. At each time step LSTM can choose to read from, write to, or reset the sell using gating mechanisms. They include four gates, generally denoted as \(i\), \(o\), \(f\), and \(\tilde{c}\), corresponding to the input, the output, the forget, and the new memory gate. In \Cref{eq:lstm1,eq:lstm2,eq:lstm3,eq:lstm4,eq:lstm5,eq:lstm6}, \(W_z\) and \(U_z\) correspond to weights of the input, \(x_t\), and the hidden state, \(h_{t-1}\), where \(z\) can either be the input gate, the output gate, and the forget gate or the memory gate, depending on the activation being calculated. \textcolor{black}{The $b_i$, $b_o$, $b_f$ and $b_c$ are the biases of the input gate, the output gate, the forget gate and the memory cell, respectively. Also, the operation $\odot$ denotes the element-wise vector product.}

\begin{figure}[H]
	\centering
	\includegraphics[width=0.61\linewidth]{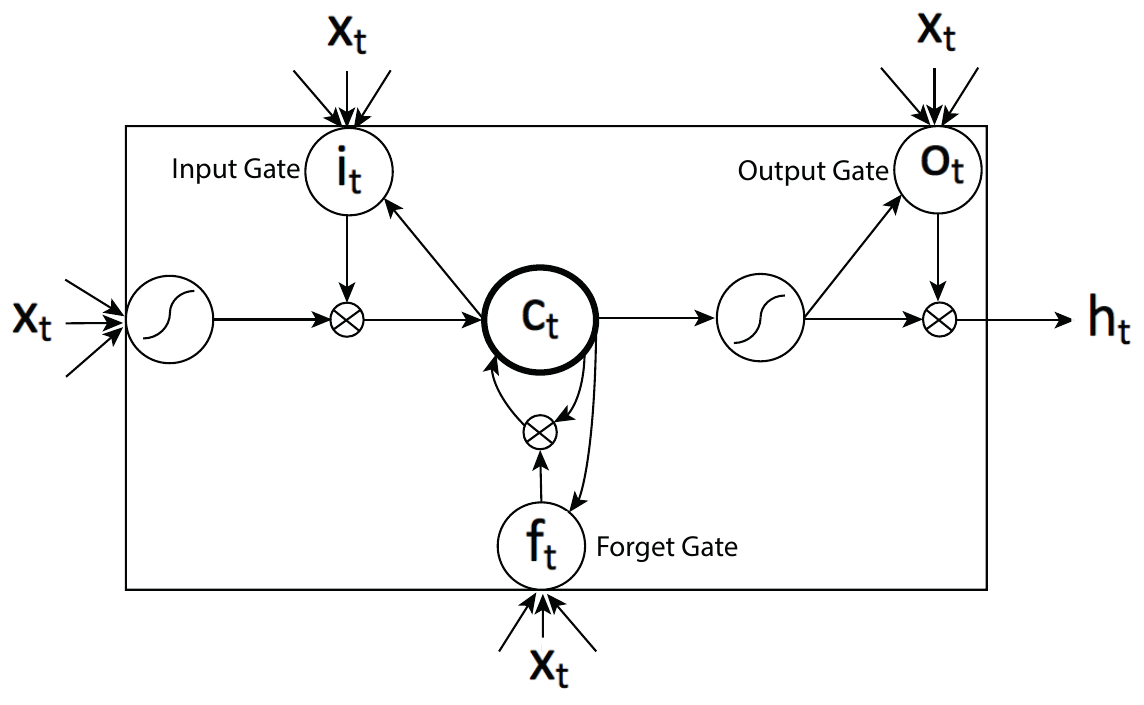}
	\caption{Long Short-Term Memory Cell}
	\label{fig:lstm}
\end{figure}
%
Intuitively,  the gate controls the flow of information to enter and exit from the cell at each time step.
In the first step, the forget gate, \(f_t\), decides how much of the previous state to take into account.
\begin{equation}
	f_t = \sigma(W_f x_t + U_f h_{t-1} + b_f) \label{eq:lstm1}
\end{equation}
Next, the input gate, \(i_t\), decides which values to update and a new memory generation stage, \(\tilde{c}_t\), creates a vector of new candidate values that could be added to the state.
\begin{equation}
	i_t = \sigma(W_i x_t + U_i h_{t-1} + b_i) \label{eq:lstm2}
\end{equation}
\begin{equation}
	\tilde{c}_t = \tanh(W_c x_t + U_c h_{t-1} + b_c) \label{eq:lstm3}
\end{equation}
Now, the new memory gate, \(c_t\) forgets the old cell state, \(\tilde{c}_{t-1}\) and gates the new memory cell, \(\tilde{c}_t\). 
\begin{equation}
	c_t = f_t \odot \tilde{c}_{t-1} + i_t \odot \tilde{c}_t \label{eq:lstm4}
\end{equation}
Finally, the output gate, \(o_t\), decides what parts of the cell state to output, then the output will be filtered by a tanh layer, to output the desired parts of it.
\begin{equation}
o_t = \sigma(W_o x_t + U_o h_{t-1} + b_o) \label{eq:lstm5}
\end{equation}
\begin{equation}
h_t = o_t \odot \tanh(c_t) \label{eq:lstm6}
\end{equation}
We built the LSTM model from an embedding layer (of dimensionality 512), an LSTM layer (with 512 network units for each gate) with dropout regularization, and finally, a sigmoid activation function applied to the output of the LSTM.
\par

Convolutional neural networks (CNNs)\cite{lecun1995convolutional}, are comprised of an input layer, one or more hidden layers, and an output layer. The hidden layers are a combination of convolutional layers, relu layers, pooling layers, and fully-connected layers. Convolutional layers will compute the output of neurons that are connected to only a small region in the input, each computing a dot product between their weights and a small region they are connected to in the input volume. Relu layers will apply an element-wise activation function which zeros out negative inputs and is represented as $\max$(0, $x$). Pooling layers will perform a down-sampling operation along the spatial dimensions (width and height) of the input. Fully-connected layers have neurons that are functionally similar to convolutional layers (compute dot products) but are different in that they are connected to all activations in the previous layer. The last fully-connected layer is called the output layer and it will compute the class scores. Stacking these layers will form a full CNN architecture as demonstrated in Figure \ref{fig:generalschemepaper}.
\\
\par
\begin{figure}
	\centering
	\includegraphics[width=0.7\linewidth]{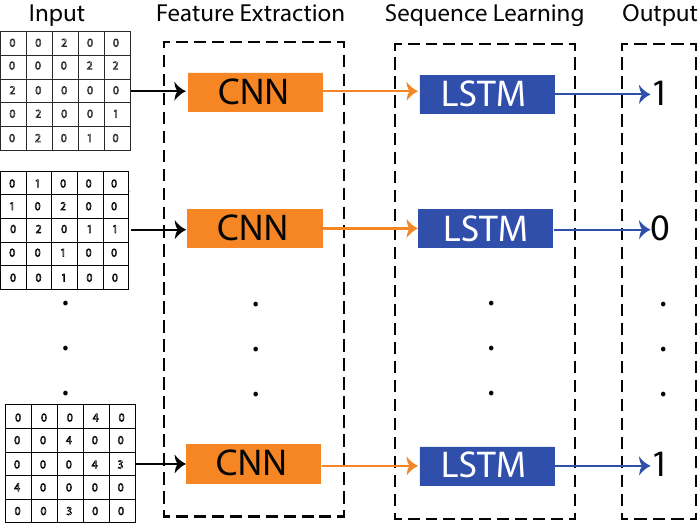}
	\caption{CNN-LSTM Architecture}
	\label{fig:cnnlstm1}
\end{figure}

CNN-LSTM is a combination of CNN for feature extraction and LSTM for summarization of the extracted features. Adopting an LSTM for aggregating the features enables the network to take the global structure into account while local features are extracted by CNN layers as represented in Figure \ref{fig:cnnlstm1}. These features can be used in various heterogeneous network mining tasks, such as \textcolor{black}{clustering \cite{liu2019multiple}, classification \cite{yu2018infrared}} and so on which we used for prediction. Our CNN-LSTM model uses an embedding layer (of dimensionality 1000), a one dimensional CNN layer of $5 \time 5$ convolutions interspersed with a $64 \time 64$ max pooling, an LSTM layer (with 1000 network units for each gate), and finally, a sigmoid activation function applied to the output of the LSTM.

\par 
One of the primary methods in the deep neural network is MLP. In this section this method is briefly  described  just for a more detailed comparison with other introduced methods \cite{ordonez2016deep}.

\textcolor{black}{The proposed method has the following properties:
\begin{enumerate}
	\item We added weights to the input layer by considering different meta-paths in heterogeneous networks.
	\item We have used deep representation learning which automatically extracts summarized features from each node in heterogeneous networks.
\end{enumerate}
}
\section{Experiments} \label{ssec:Results}
We evaluate the proposed approach based on  benchmark datasets and standard evaluation measures which are introduced in the  following subsections. Then, the performance of the proposed approach is described in comparison with the earlier well-known methods.
For topic diffusion and cascade prediction, initially, we need to select a topic like ``data mining''. The authors who had a paper in the field of related topic considered as active. After that, we should predict the activation of other inactive nodes in the next timestamps with the proposed methods.  
\subsection{Datasets}
The \emph{DBLP}, \emph{PubMed}, and \emph{ACM} are benchmark real network datasets which are employed in our experimental studies. 
\begin{itemize}
	\item \emph{DBLP}: This dataset is about computer science bibliography among authors in main conferences and publications \cite{Dataset}. Objects indicate authors in this network. Different meta-path such as APA (Author-Paper-Author), ACA (Author-Conference-Author), APAPA (Author-Paper-Author-Paper-Author), and ACACA (Author- Conference -Author- Conference -Author) are considered. Different topics are extracted from this dataset, and information diffusion about a specific topic is investigated. This dataset contains information from 1954 to 2016.
	\item \emph{PubMed}:  This dataset consists of medical science bibliography among authors in main conferences and publications in this domain\cite{DataRep}. In this network, the authors are represented by objects and meta-paths APA and APAPA are used. The dataset contains bibliographic information from 1994 to 2003. 
	\item \emph{ACM}: This dataset consists of the bibliographic information of publications of the Association for Computing Machinery (ACM) \cite{Dataset}. In this network, we used APA and APAPA meta-paths. The dataset contains information from 1959 to 2009. 
	\textcolor{black}{\item \emph{APS}: This dataset consists of the all American Physical Society (APS) journal articles \cite{APS}. In this network, we used APA and APAPA meta-paths. The dataset contains information from 1913 to 2016.}
	\textcolor{black}{\item \emph{Citeseer}: This citation network extracted from the CiteSeer digital library. CiteSeer is an evolving scientific literature digital library and search engine that has focused primarily on the literature in computer and information science \cite{Citeseer}. Nodes are publications and the directed edges denote citations. We used 50,000 nodes of this dataset.}
\end{itemize}

A summary about the datasets are given in Table \ref{tab:table1}.
\begin{table}[t!]
	\caption {The main properties of applied datasets}
	\begin{center}
		\begin{tabular}{ |c|c|c|c| } 
			\hline
			\textbf{Dataset} & \textbf{Authors} & \textbf{Papers} \\
			\hline
			\emph{DBLP} & 215222 & 104940 \\ 
			\hline
			\emph{PubMed}& 7140093 & 4271136 \\ 
			\hline
			\emph{ACM}& 468114 & 649526 \\ 
			\hline
			\emph{APS}& 96644 & 96060 \\ 
			\hline
			\emph{Citeseer}& -- & 50000 \\ 
			\hline			
		\end{tabular}
	\end{center}
	\label{tab:table1}
\end{table}

\subsection{Evaluation Measures}
For topic diffusion, we use Precision and Recall criteria to assess the performance. These measures are defined as,
\begin{equation} \label{eqTf}
{Precision=\frac{TF}{TF+FP}, Recall=\frac{TP}{TP+FN}}
\end{equation}

{Where True Positive (TP) is the active nodes that are correctly tagged as active by the algorithm, True Negative (TN) is the inactive nodes that are correctly tagged as inactive by the algorithm, False Positive (FP) is the active nodes that are falsely tagged as inactive by the algorithm, and False Negative (FN) is the inactive nodes that are falsely tagged as active by the algorithm. }\\
Furthermore, AUPR( Aurea under Precision Recall) curve  is used by plotting Precision against Recall. The higher they are, the better the model is.\par
The MSE(Mean Squared Error) and AP(Average Precision) is applied to measure the accuracy of cascade prediction task. AP summarizes a precision-recall curve as the weighted mean of precisions achieved at each threshold \cite{zhu2004recall}.\par
On training and test data selection, we first consider all nodes with published papers as our particular topic of interest as active ones and the rest nodes as inactive. 
At time \(t\), the training and test sets are selected as follows:\\
\textit{Training set:} Those within the time period from \textit{\(t-4\)} to \textit{\(t-1\)}  and from \textit{\(t-4\)} to \textit{\(t\)} are considered as the training set for topic diffusion and cascade prediction separately.\\
\textit{Test set:} Those within the time period from \textit{\(t-1\)} to \textit{\(t\)} are considered as the test set in topic diffusion. Additionally, the nodes tagged as active up to the time \(t-1\) are considered as the seed nodes that are activated initially in the start of the diffusion process. In the case of cascade prediction, the sequence of nodes at  time \textit{\(t\)} are employed as the test set.

\subsection{Performance Evaluation}
The proposed approach \emph{HDD} is applied on topic diffusion and cascade prediction in comparison with the earlier well-known methods. 
On topic diffusion on heterogeneous networks,  the performance of \emph{HDD} is compared to \emph{MLTM-R} which is the only related  work in this category. \textcolor{black}{The \emph{Planetoid}\cite{yang2016revisiting} is one of the well-known embedding methods that is employed for our comparison.}
On cascade prediction, we employ \emph{Node2vec} and \emph{DeepCas} as well-known feature learning techniques that are presented in the following:
\begin{enumerate}
	\item \textcolor{black}{Planetoid\cite{yang2016revisiting}: It presents a learning frame-work based on graph embeddings which trains an embedding for each instance to jointly predict the class label and the neighborhood context in the graph.}
	\item Node2vec\cite{grover2016node2vec}: It learns a mapping of nodes to a low-dimensional space of features that maximizes the likelihood of preserving network neighborhoods of nodes. This method is selected as a representative work in the  node embedding methods.     
	\item DeepCas\cite{li2017deepcas}: It is initiated on paths samples from different snapshots of a graph. They used a GRU network to transform  path samples into a single vector.\\
	\item \textcolor{black}{DeepWalk\cite{perozzi2014deepwalk}: Inspired by the Word2Vec method \cite{mikolov2013distributed}, Perozzi et al. \cite{perozzi2014deepwalk} proposed the DeepWalk that generates random paths over a graph. It learns the new node representation by maximizing the co-occurrence probability of the neighbors in the walk.}
\end{enumerate}


\subsubsection{ACM dataset}
\begin{figure}[H]
	\centering
	\includegraphics[width=1\linewidth,height=0.45\textwidth]{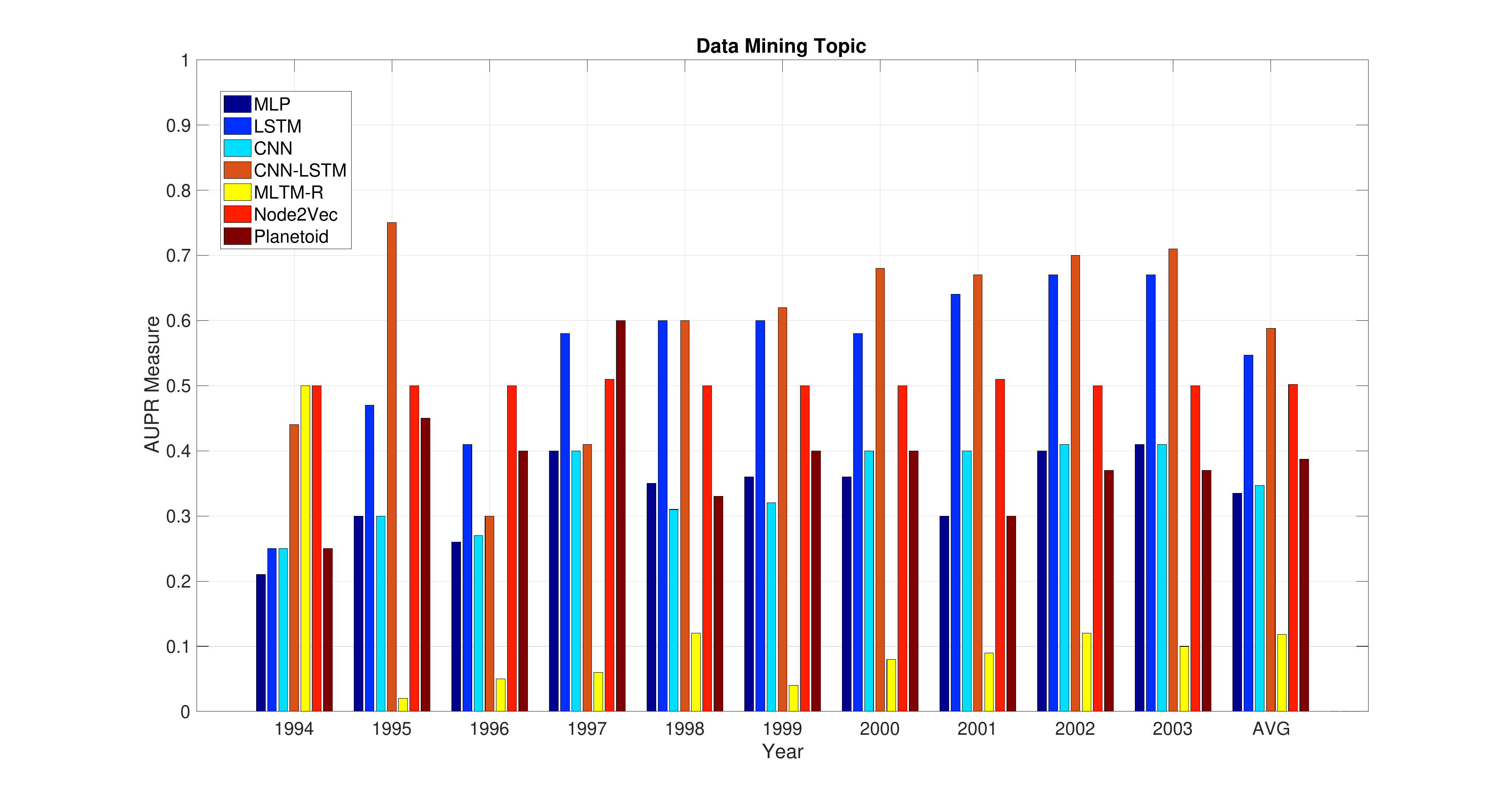}
	\caption{AUPR measure in ACM dataset for Data Mining topic (1994- 2003)}
	\label{fig:ACMdatamining}
\end{figure}

\begin{figure}[H]
	\centering
	\includegraphics[width=1\linewidth,height=0.45\textwidth]{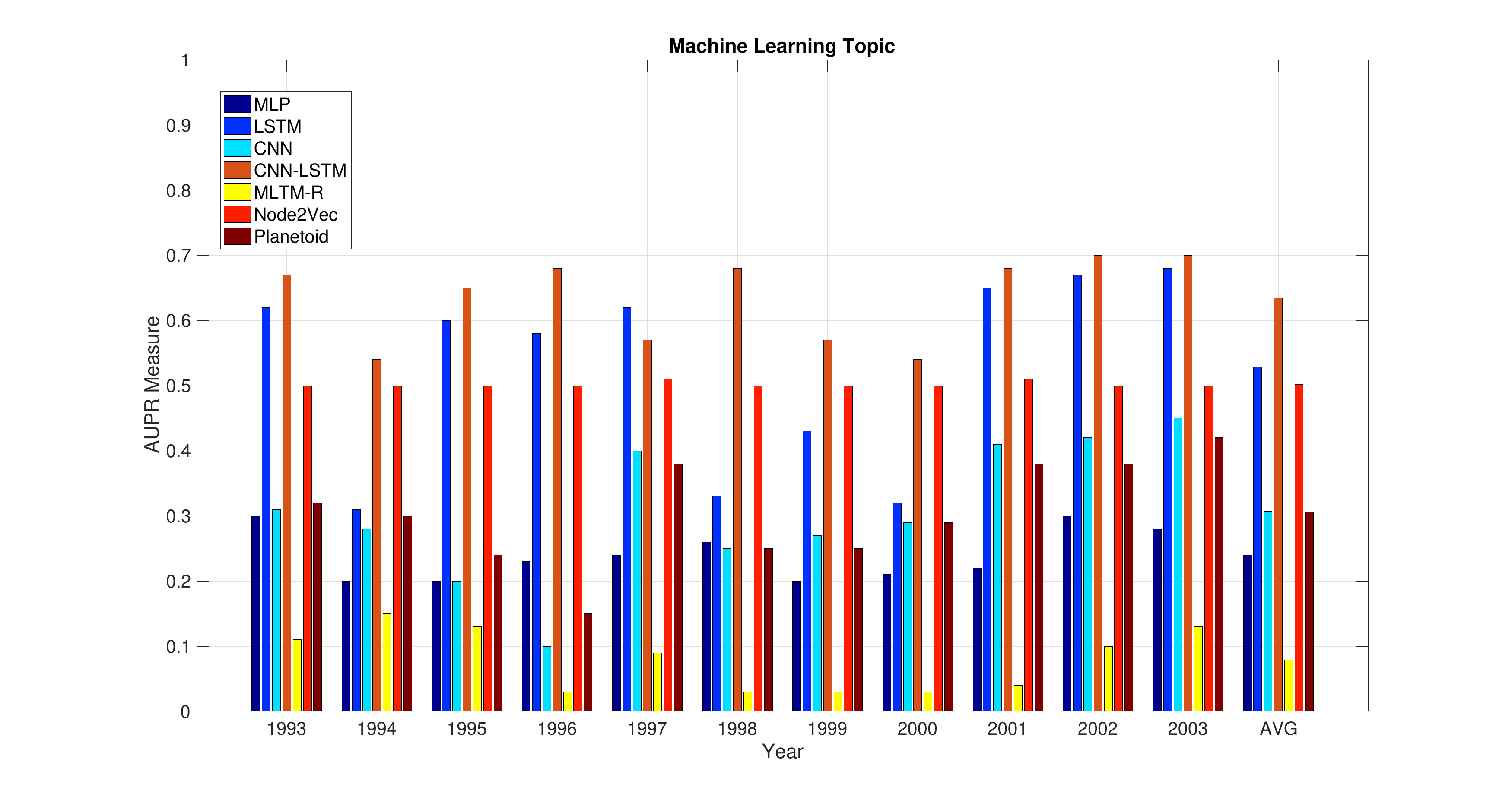}
	\caption{AUPR measure in ACM dataset  for Machine Learning topic (1993- 2003)}
	\label{fig:ACMmachinelearning}
\end{figure}

\begin{figure}[H]
	\centering
	\includegraphics[width=1\linewidth,height=0.45\textwidth]{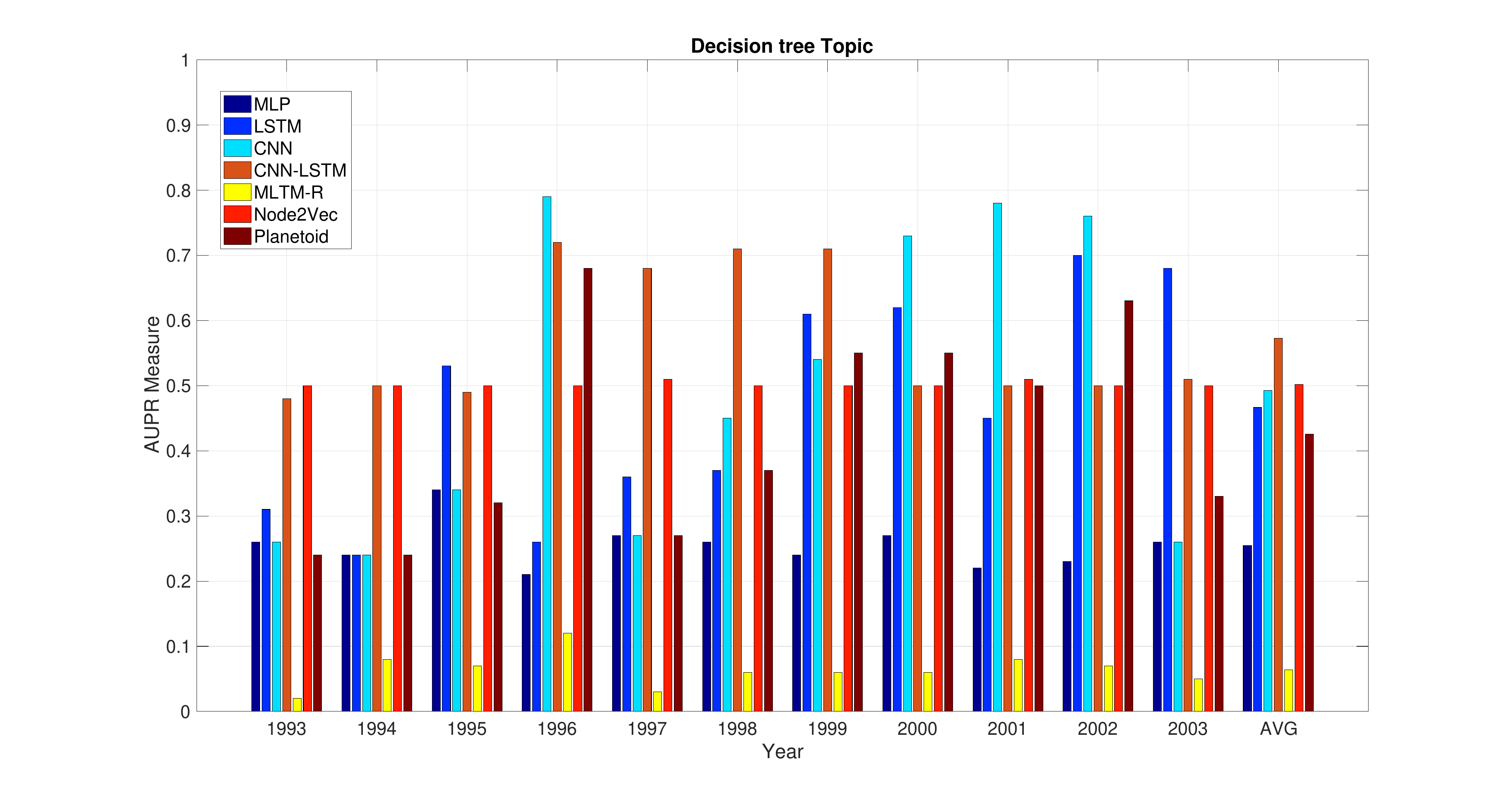}
	\caption{AUPR measure in ACM dataset for Decision tree topic (1993- 2003)}
	\label{fig:ACMDesicionTree}
\end{figure}

In this data set, the topics \emph{``Data Mining''}, \emph{``Machine Learning''} and \emph{``Decision Tree''} are selected due to having the ground-truth about them in this dataset. 
Figures \ref{fig:ACMdatamining}, \ref{fig:ACMmachinelearning} and \ref{fig:ACMDesicionTree} represent topic diffusion results on \emph{``Data Mining''},  \emph{``Machine Learning''} and \emph{``Decision Tree''} topics on \emph{ACM}. \textcolor{black}{It can be observed that the LSTM and CNN-LSTM have a significant improvement rather than the other methods. In the dataset, LSTM increased the AUPR measure around 35\% and 4\% rather than MLTM-R and Node2Vec respectively and CNN-LSTM enhanced the AUPR up to 50\% and 10\% compared to MLTM-R and Node2Vec methods.  After these methods, CNN is better than the MLTM-R and MLP which raised the AUPR by 30\% in comparison with MLTM-R method. }

\subsubsection{DBLP dataset}
\begin{figure}[H]
	\centering
	\includegraphics[width=1\linewidth,height=0.45\textwidth]{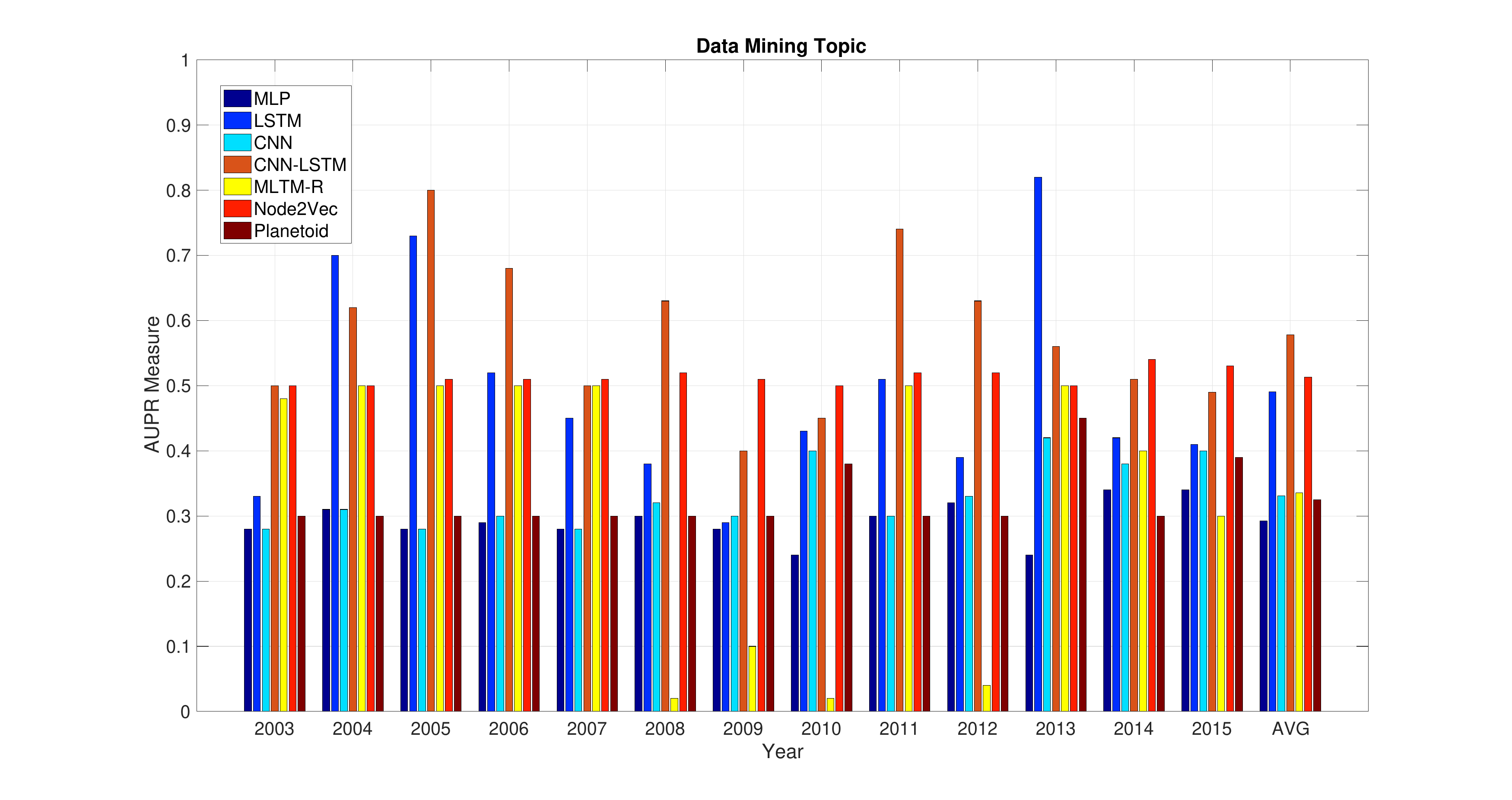}
	\caption{AUPR measure in DBLP dataset for Data Mining topic (2003- 2015)}
	\label{fig:DBLPdatamining}
\end{figure}

\begin{figure}[H]
	\centering
	\includegraphics[width=1\linewidth,height=0.45\textwidth]{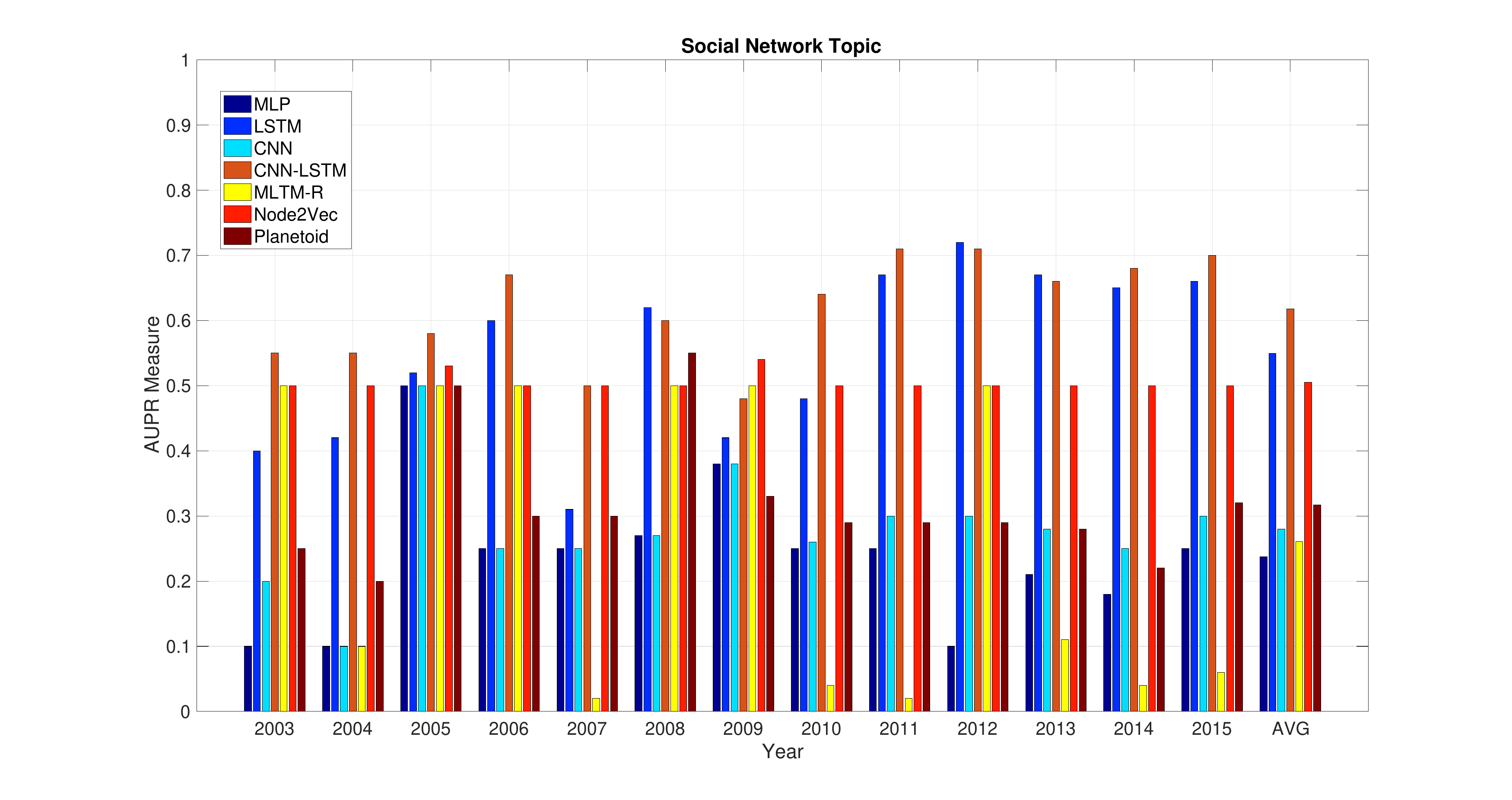}
	\caption{AUPR measure in DBLP dataset for Social Network topic (2003- 2015)}
	\label{fig:DBLPSocialNetwork}
\end{figure}

\begin{figure}[H]
	\centering
	\includegraphics[width=1\linewidth,height=0.45\textwidth]{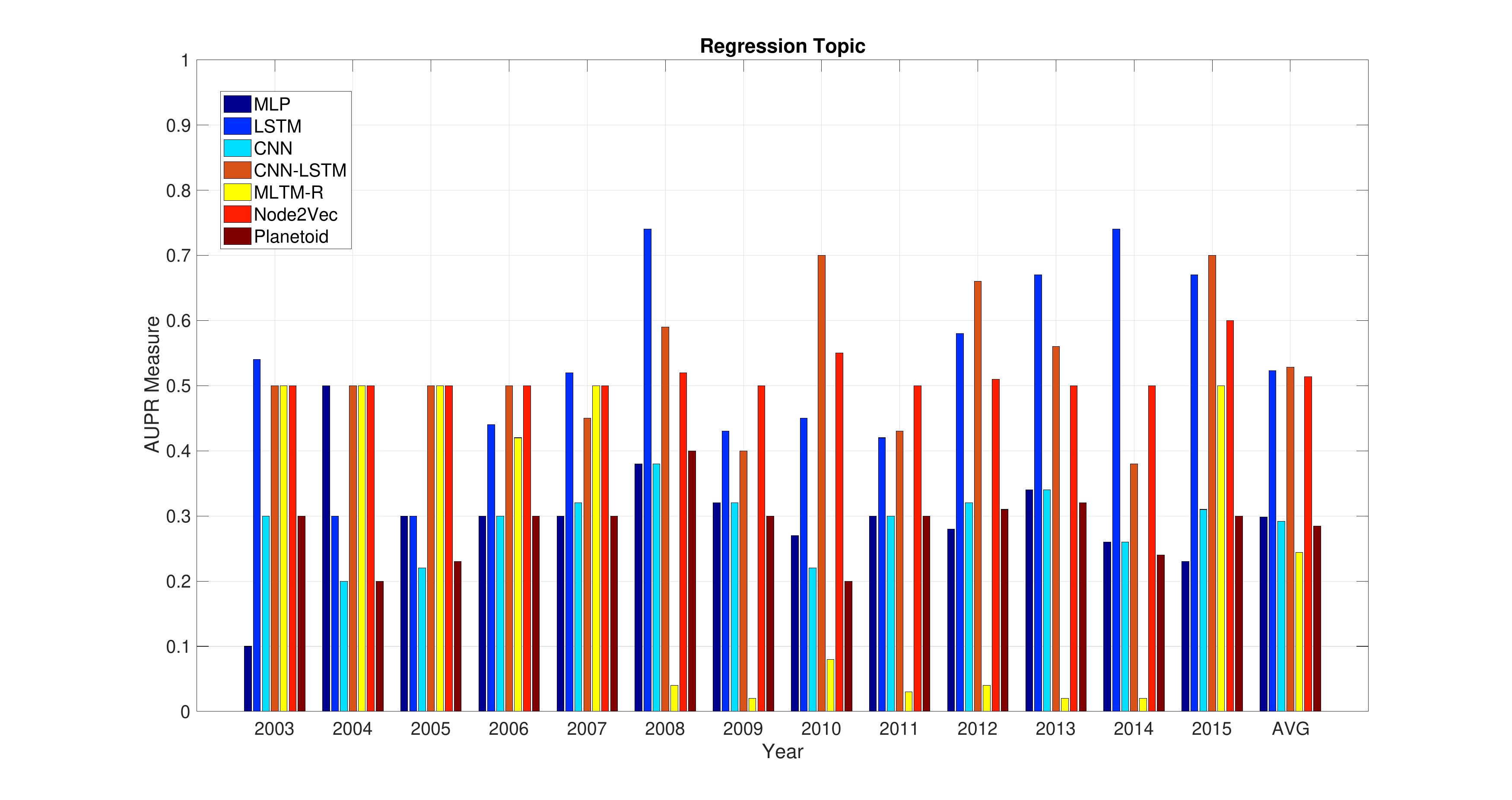}
	\caption{AUPR measure in DBLP dataset for Regression topic (2003- 2015)}
	\label{fig:DBLPRegression}
\end{figure}

These  topics \textit{``Data mining''}, \textit{``Social Network''} and \textit{``Regression''} were selected in DBLP. We can infer from Figures \ref{fig:DBLPdatamining}, \ref{fig:DBLPSocialNetwork} and \ref{fig:DBLPRegression} that LSTM and CNN-LSTM still have better results but here, versus ACM dataset, MLTM-R method on average shows a slight growth compared with MLP. The results comparison showed that CNN-LSTM returned high AUPR (overall AUPR about 60\%) followed by LSTM and CNN.

\subsubsection{PubMed dataset}
\begin{figure}[H]
	\centering             
	\includegraphics[width=1\linewidth,height=0.45\textwidth]{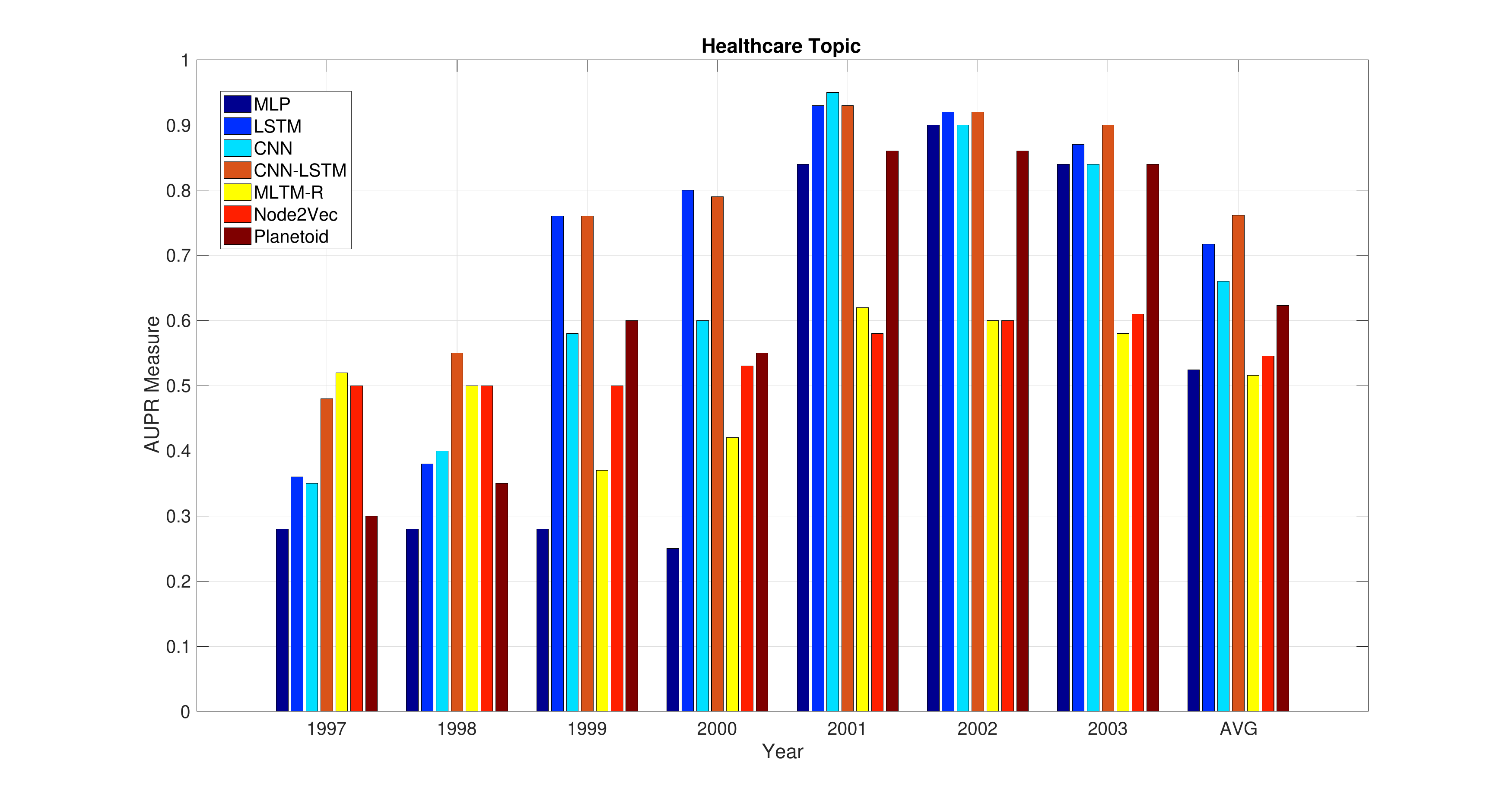}
	\caption{AUPR measure in PubMed dataset for Health care topic (1997- 2003)}
	\label{fig:PubMedRegression}
\end{figure}

In this dataset \textit{``Health care''} topic was examined. As shown in the figure \ref{fig:PubMedRegression}, LSTM and CNN-LSTM improved the AUPR Measure in comparison with other used methods. Here, CNN makes the results a little better against MLP  due to the volume of data. According to AUPR, CNN-LSTM and LSTM outperformed MLTM-R by 25\% and 18\%, Node2Vec by 21\% and 17\% and \textcolor{black}{Planetoid by 13\% and 9\% respectively.}

\subsubsection{APS Dataset}
\begin{figure}[H]
	\centering
	\includegraphics[width=1\linewidth]{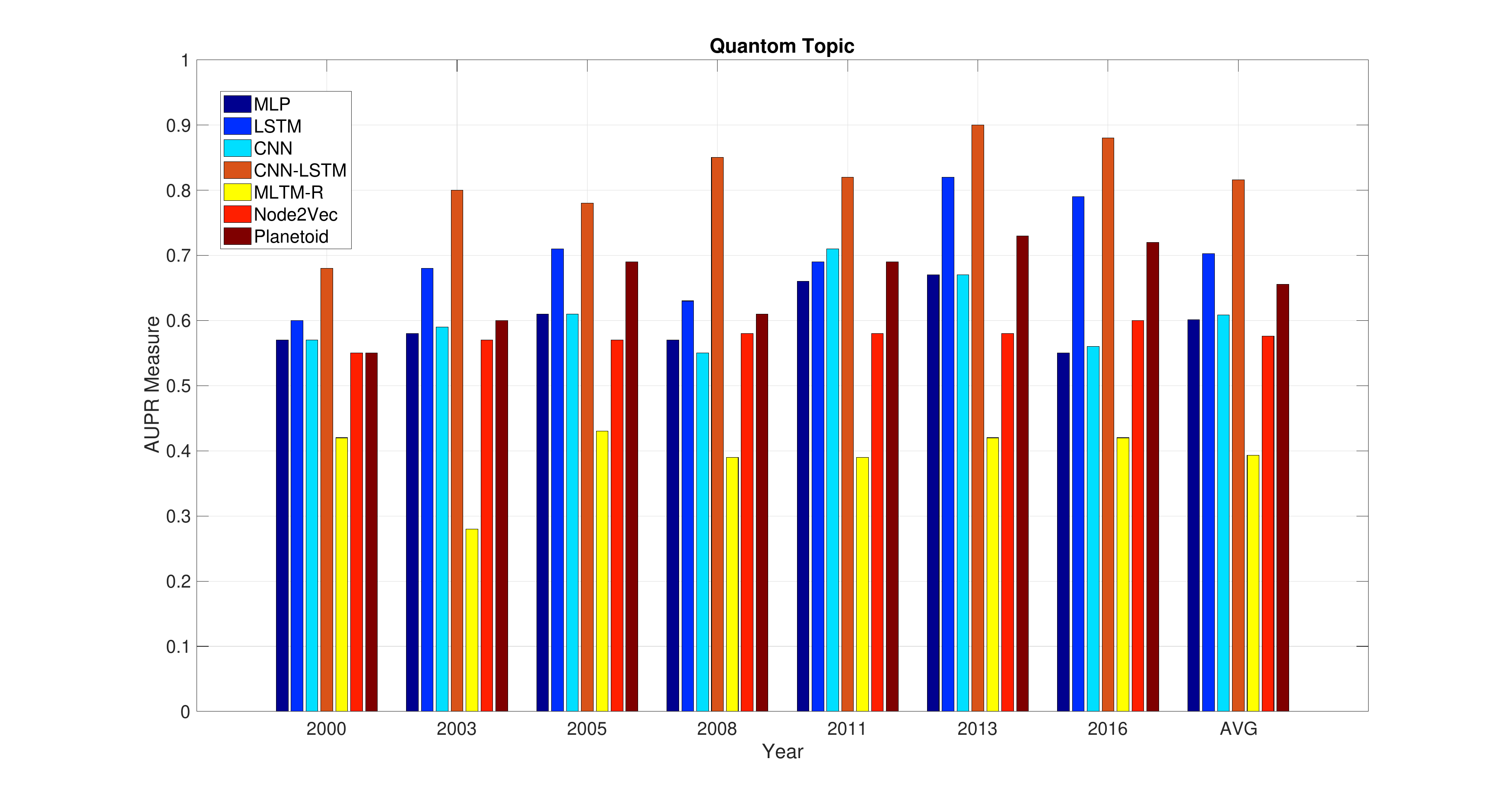}
	\caption{AUPR measure in Citeseer dataset for Quantum topic (2000- 2016)}
	\label{fig:quantom}
\end{figure}

\textcolor{black}{We used \textit{``Quantum''} topic for this dataset. Figure \ref{fig:quantom} identifies further improvement (roughly 42\%) of CNN-LSTM method against MLTM-R method and after this method, LSTM is highest. CNN-LSTM and LSTM also increase the AUPR against Node2Vec about 0.24\% and 12\% and Planetoid about 16\% and 4\% in order.}

\subsection{Cascade Prediction}
\begin{figure}[H]
	\centering
	\vspace{-0.2cm}
	\subfigure[{MSE error for Data Mining topic}]{\includegraphics[width=0.46\textwidth, height=0.36\textwidth]{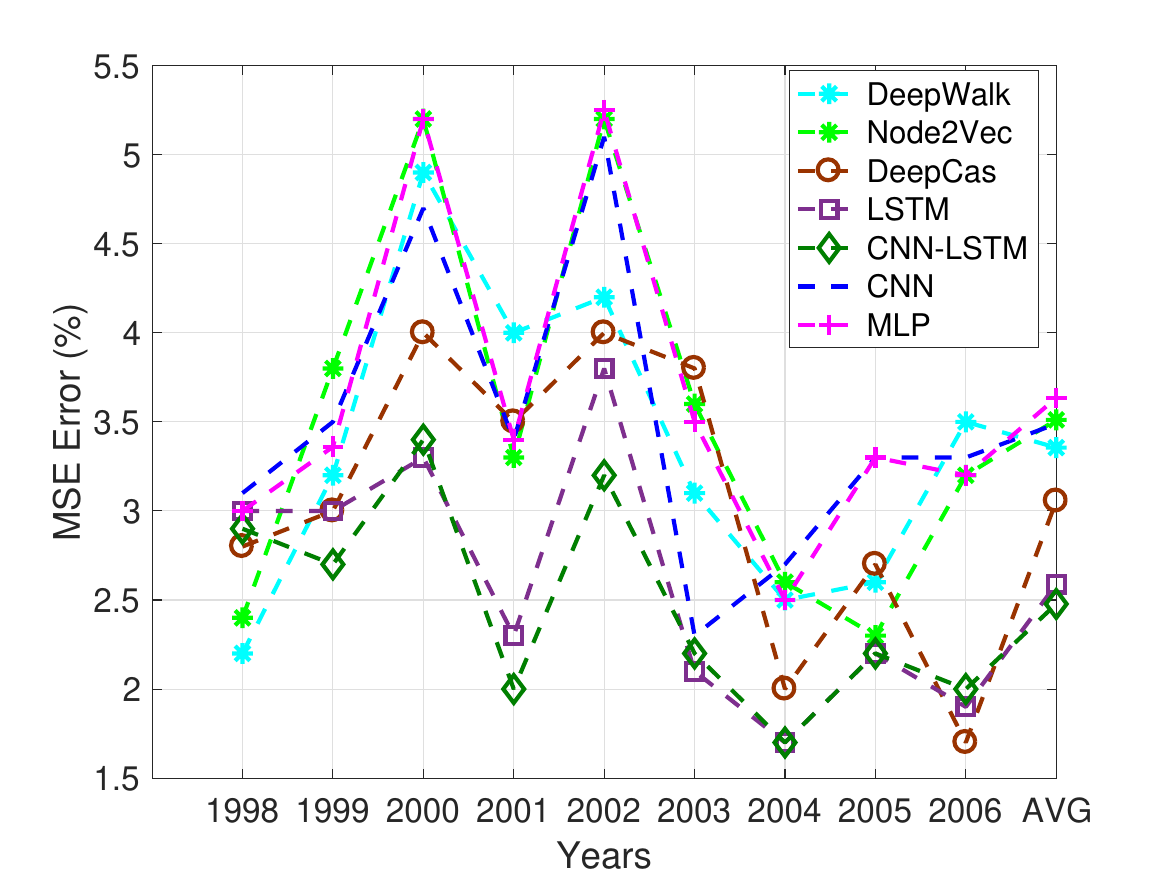}
		\label{fig:msedataminingcascade}}
	\subfigure[AP measure for Data Mining topic]{\includegraphics[width=0.46\textwidth, height=0.36\textwidth]{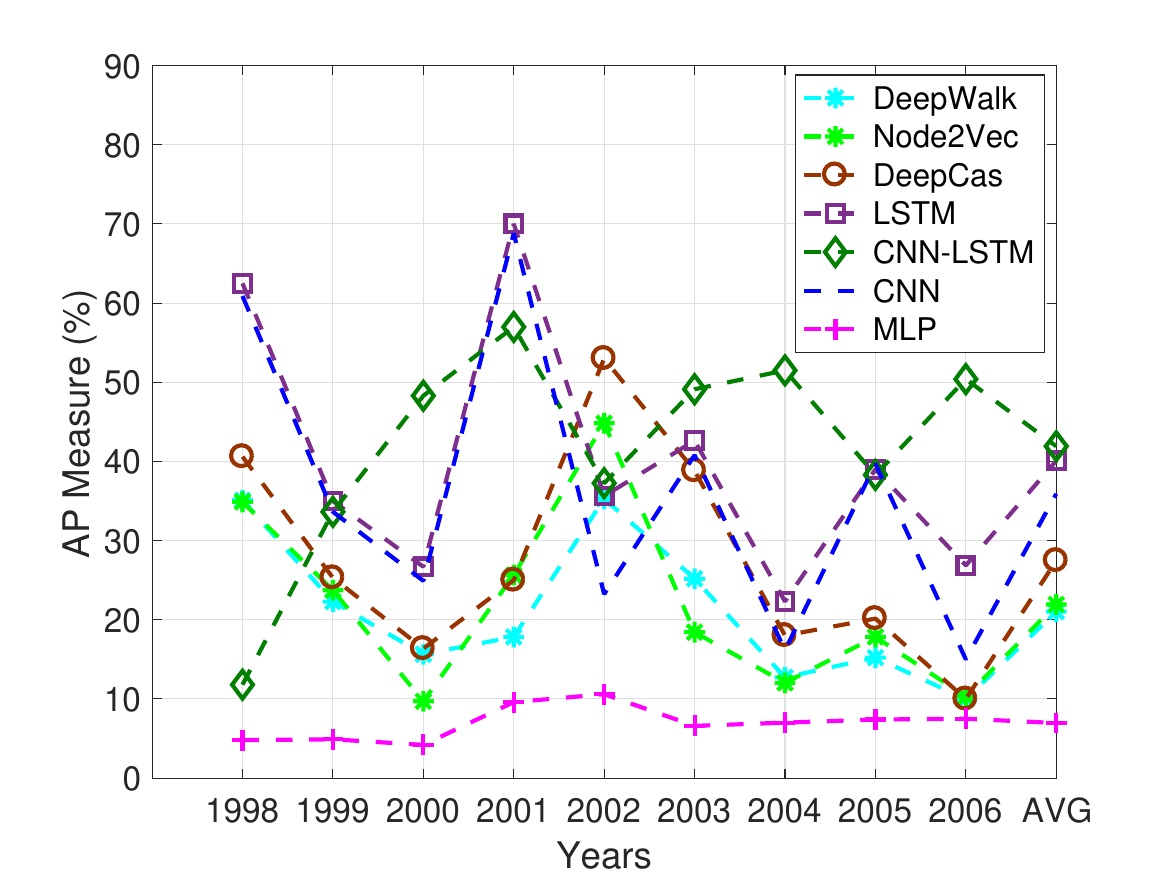}
		\label{fig:msedataminingcascade1}}
	\vspace{-0.3cm}
	\caption{AP Measure and MSE error for Data Mining topic}
	\label{MAPDatamining}
\end{figure}

\begin{figure}[H]
	\centering
	\vspace{-0.2cm}
	\subfigure[{MSE error for Machine Learning topic}]{\includegraphics[width=0.46\textwidth, height=0.36\textwidth]{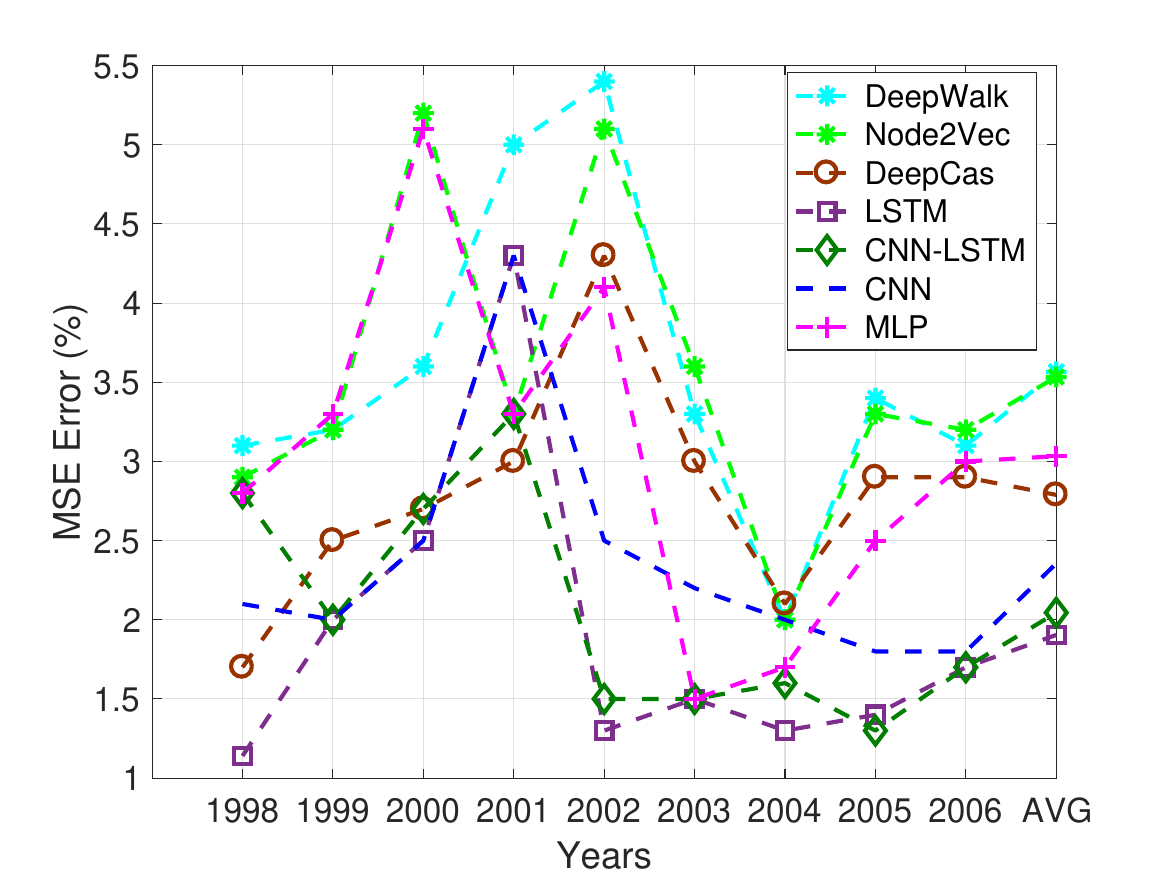}
		\label{fig:msemachinelearningcascade}}
	\subfigure[AP measure for Machine Learning topic]{\includegraphics[width=0.46\textwidth, height=0.36\textwidth]{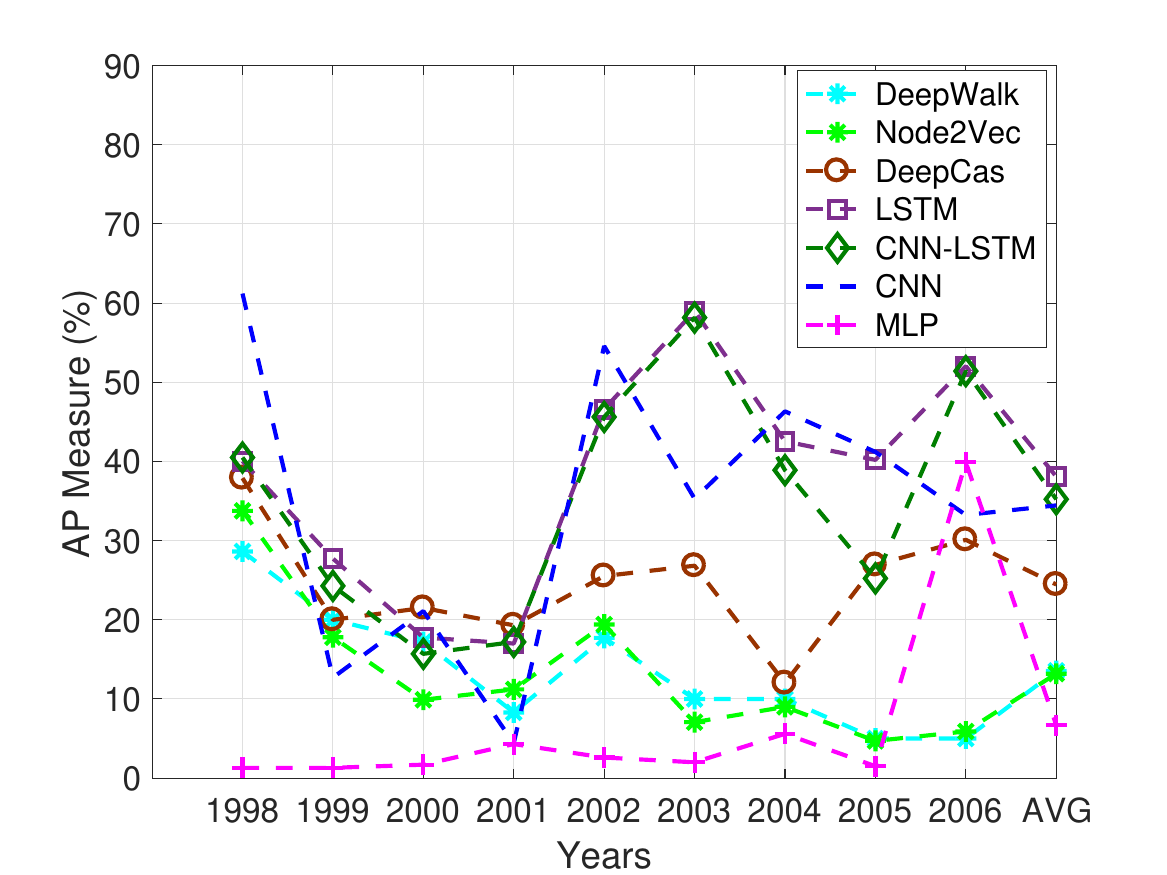}
		\label{fig:msemachinelearningcascade1}}
	\vspace{-0.3cm}
	\caption{AP Measure and MSE error for Machine Learning topic}
	\label{MAPMAchine learning}
\end{figure}

We evaluate the prediction of cascades through the DBLP and Citeseer datasets because of existing citation relations. In DBLP dataset, we created cascades output according to PCP(paper-citation-paper) in data mining, machine learning topics and construct the meta-paths graph input based on PCP(paper-citation-paper), PVP(paper-conference-paper). As shown in Figures \ref{MAPDatamining} and \ref{MAPMAchine learning}, LSTM and CNN-LSTM outperform the other methods by a significant margin.

\begin{figure}[H]
	\centering
	\includegraphics[width=1\linewidth]{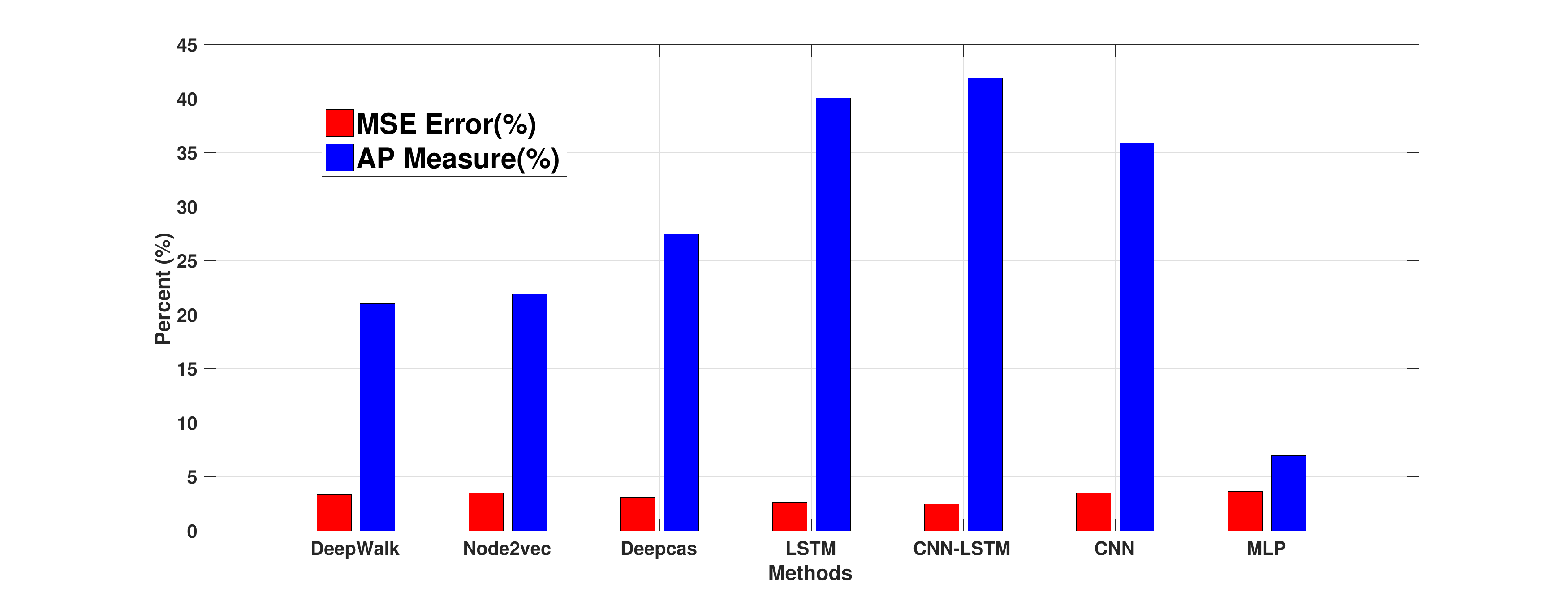}
	\caption{AP Measure and MSE error for Citeseer Dataset}
	\label{fig:citeseercascade}
\end{figure}

\textcolor{black}{The Citeseer dataset is used in our experiments by considering  PCP(paper-citation-paper) and PCPCP(paper-citation-paper-citation-paper) meta-paths. The results show the superiority of LSTM and CNN-LSTM in comparison with the other methods as represented in Figure \ref{fig:citeseercascade}}.

\textcolor{black}{The differences in results for each dataset indicate the importance  of  some main characteristics such as the topological structure, the context, and  frequency of  the specified topic in each network dataset. For example, the increasing pattern in PubMed results on topic diffusion is considerable due to the amount of topics as well as the number of nodes.}

\section{Conclusion}\label{ssec:Conclusion}
This paper studied topic diffusion and cascade prediction in heterogeneous networks. With this aim, we introduced a new end to end graph representation learning method in heterogeneous networks. The end-to-end predictor, HDD, outperformed the feature-based machine learning methods and alternative author embedding and meta-path graph embedding methods.  The HDD model captures the influence of authors and cascades  in different timestamps and meta-paths. Besides, employing  the entire meta-paths through deep structures as an alternative to the ad-hoc based relations  can significantly improve the prediction performance. In addition, the HDD model is flexible to use in other multilayer networks. We demonstrated the advantages of deep architecture methods in real-world networks such as DBLP, PubMed and ACM. Our experimental results on three real data sets verified the effectiveness and efficiency of our methods, LSTM and CNN-LSTM. 
The performance of the proposed method was compared with the state-of-the-art techniques. The obtained results showed that the proposed method outperforms the earlier ones.  As future work, we are interested in a combination of graph summarization and deep learning to improve the results.

\end{document}